\documentclass[12pt]{article}
\usepackage{amsmath,amssymb,booktabs}

\usepackage{cancel}

\makeatletter

\@addtoreset{equation}{section}
\makeatother

\usepackage[pdftex]{graphicx,color}
\usepackage{epstopdf}
\usepackage[bookmarks=true,bookmarksnumbered=true,bookmarkstype=toc]{hyperref} 

\usepackage{color}

\definecolor{green}{cmyk}{1,0,1,0}

\usepackage{ulem}  

\begin{document}

\title{\vbox{
\baselineskip 14pt
\hfill \hbox{\normalsize KYUSHU-HET-161}\\
\hfill \hbox{\normalsize KYUSHU-RCAPP-2016-01}\\
} \vskip 1.7cm
TeV scale mirage mediation in NMSSM and precision Higgs measurement
\footnote{Talk presented at the International Workshop on Future Liner Colliders (LCWS15), Whistler, Canada, 2-6 November 2015 based on the collaboration in \cite{Hagimoto:2015tua}}
 \bf  \vskip 0.5cm
}
\author{%
Ken-ichi~Okumura
\\*[20pt]
{\it \normalsize 
Department of Physics, Kyushu University, Fukuoka 819-0395,
Japan} \\
}

\date{}

\maketitle
\thispagestyle{empty}

\begin{abstract}
We propose that the TeV scale mirage mediation in the Next-to-Minimal Supersymmetric Standard Model (NMSSM) is a novel solution for the little hierarchy problem in supersymmetry. The 125 GeV Higgs boson can be accommodated with the new quartic coupling or the mixing with the singlet.  The fine-tuning measures are estimated numerically and found as low as 10\% or below for $1.5$ TeV gluino and $1$ TeV stop. 
The higgsino can be as heavy as 500 GeV without deteriorating the fine-tuning.
An undesirable singlet-doublet mixing in the Higgs sector is suppressed due to approximate scale symmetries.
We investigate the couplings of the Higgs bosons and discuss the prospects for studying them at LHC and ILC. 

\end{abstract}

\newpage

\setcounter{footnote}{0}

\section{Introduction}

Supersymmetry (SUSY) is a promising candidate for physics beyond the Standard Model (SM). It stabilizes the big hierarchy between the Planck scale and the electroweak scale (1 part in $10^{32}$ for the squared mass scale). Furthermore, the minimal supersymmetric extension of the SM (MSSM) predicts a precise gauge coupling unification. It is also equipped with a natural candidate of the dark matter. In 2012, ATLAS and CMS collaborations discovered the Higgs boson, the last missing piece of the SM \cite{Aad:2012tfa, Chatrchyan:2012ufa}. Its mass is 125 GeV \cite{Aad:2015zhl}, which means that the self-coupling of the Higgs boson is perturbative. This perfectly matches with the prediction of the minimal SUSY. 

SUSY is such a wonderful theory with full of conceptual successes. However, now it is facing with an annoying 'little fine-tuning problem'.  
LHC Run I has finished and it did not find any sign of the colored SUSY particles. The lower bound of the mass of the gluino is now roughly 1.4 TeV. Those of the first/second generation squarks and the third generation squarks are about 1 TeV and 700 GeV in simplified models (for example see, \cite{Ulmer:2016ljv}). 
Among them, the masses of the gluino and stops are mixed up with $m_{H_u}^2$, the SUSY breaking mass parameter of the Higgs boson, via the renormalization group evolution. Without fine-tuning the initial conditions, they are expected to be similar in their sizes.
While in the MSSM, the Z boson mass, $m_Z$ which represents the scale of the electroweak breaking is determined as,
\begin{equation} 
\frac{m_Z^2}{2} \simeq -m_{H_u}^2 -|\mu|^2 +\mathcal{O}\left(\frac{1}{\tan^2\beta}\right), 
\label{eq:fine-tuning}
\end{equation} 
with $1/\tan\beta$ expansion. Here $\mu$ is the SUSY preserving higgsino mass whose origin is in most cases different from $m_{H_u}^2$.
This means that $m_Z=91$ GeV is the result of a fine-tuning between $m_{H_u}^2$ and $\mu^2$ worse than 1\%.
A possible solution of this problem is a SUSY breaking mediation mechanism which, with some underlying reasons, exactly realizes $m_{H_u}^2=\mathcal{O}(m_Z^2)$ at the electroweak scale, simultaneously with the gluino and stops as heavy as $\mathcal{O}(1)$ TeV or more (little SUSY hierarchy). TeV scale mirage mediation is proposed as one of such mechanisms \cite{Choi:2005hd, Choi:2006xb}\cite{Kitano:2005wc}. In the MSSM, it alleviates the problem considerably, however, suffers from several problems, in particular after the discovery of the Higgs boson.
In this talk, we argue that the TeV scale mirage mediation in the Next-to-Minimal SUSY SM (NMSSM) provides a novel solution for this little SUSY hierarchy problem and discuss its phenomenology focusing on the precision Higgs measurement \cite{Kobayashi:2012ee, Hagimoto:2015tua}\footnote{Also see \cite{Asano:2012sv} as a related work.}.

\section{TeV scale mirage mediation in NMSSM}

The mirage mediation is a SUSY breaking mediation mechanism based on supergravity \cite{Choi:2004sx}\cite{Choi:2005uz}\cite{Endo:2005uy}.
It is inspired by the KKLT flux string compactification \cite{Kachru:2003aw}, however, not necessarily assumes strict string realizations. 
It is a combination of well-known two mechanisms, the modulus mediation \cite{Kaplunovsky:1993rd} and the anomaly mediation \cite{Randall:1998uk}. The order parameter of the former is the modulus mediated gaugino mass, $M_0$ and that of the latter is the gravitino mass $m_{3/2}$. 
The modulus mediation is a tree-level effect in supergravity, while the anomaly mediation works at loop-level. Thus, the latter is negligible if $M_0\sim m_{3/2}$. However, the KKLT type modulus stabilization predicts $M_0 \sim m_{3/2}/(4\pi)^2$ and both of them give the leading effect \cite{Choi:2004sx}.
The important feature of the mirage mediation is the mirage unification \cite{Choi:2005uz}.
The renormalization group corrections to the modulus mediated SUSY breaking cancel with the anomaly mediation at the mirage unification scale, $M_{mir}=M_{GUT}/(M_{pl}/m_{3/2})^{\alpha/2}$ at one-loop level.
Then, the boundary conditions for the modulus mediation appear at $M_{mir}$, not  at the unification scale.
$\alpha$ is essentially given by the ratio of $M_0$ and $m_{3/2}$, $\alpha = m_{3/2}/M_0\ln(M_{pl}/m_{3/2})$. It is determined by the scaling dimension of the uplifting potential against the modulus field in supergravity. Considering its underlying geometrical origin, $\alpha$ is not arbitrary but typically given by a ratio of small integers.
If $\alpha=2$, the boundary conditions for the modulus mediation appear around the electroweak scale. 
They are given by, 
\begin{equation}
M_{1/2} = M_0, \qquad m_i^2 = c_i M_0^2, \qquad A_{ijk} = (c_i + c_j + c_k) M_0, 
\end{equation}
with the soft SUSY breaking Lagrangian,
\begin{equation}
-{\cal L}_{Soft} = \frac{1}{2} M_{1/2} \bar{\lambda^a} \lambda^a + m_i^2 |\phi_i|^2 
+ \left\{ \frac{1}{3!}Y_{ijk} A_{ijk} \phi_i \phi_j \phi_k + {\rm h.c.}\right\}, 
\end{equation}
where $M_{1/2}$ is the mass of the gaugino $\lambda^a$, while $m_i$ is the mass of the sfermion $\phi_i$. $A_{ijk}$ is the $A$ term for the Yukawa coupling constant, $Y_{ijk}$\footnote{Here we neglect the possibility of the flavor mixing among the sfermions}. $c_i$ is the modular weight for $\phi_i$ and given by a ratio of small integers again reflecting a geometrical origin of $\phi_i$. In addition, we also have uncontrollable threshold corrections depending on the detail of the UV theory \cite{Choi:2008hn}.
For the cancellation of the renormalization group running, a sum rule, $c_i+c_j+c_k=1$ is required for non-negligible Yukawa coupling constants, $Y_{ijk}\sim {\mathcal{O}(1)}$.
If the UV theory chooses $c_{H_u}=0$, we obtain the desired little hierarchy at the electroweak scale up to the threshold corrections and 2-loop effects in the renormalization group running. This is the TeV scale mirage mediation scenario.

TeV scale mirage mediation can considerably ameliorate the little SUSY hierarchy problem, however, it has two problems in the MSSM.
The first problem is the $B\mu$ problem inherent with the model based on the anomaly mediation. The $B$ term which appears in the Higgs potential has a tree-level contribution of the order of $m_{3/2} >> M_0$. Thus, we need a delicate cancellation with another contribution to realize $B\sim M_0$ \cite{Choi:2005uz}\cite{Nakamura:2008ey}.
The second problem is the Higgs boson mass. 
It is difficult to achieve the 125 GeV Higgs boson with $M_0 \simeq 1$ TeV because the A-term for the top Yukawa coupling is fixed by the sum rule.   
The extension to the NMSSM can solve these two problems.

In the NMSSM, we introduce a singlet supermultiplet $S$ to the MSSM \cite{Fayet:1974pd}\cite{Ellwanger:2009dp, Maniatis:2009re}. 
The superpotential for the NMSSM is given by,
\begin{equation}
{\cal W} = -\lambda S H_d H_u + \frac{1}{3}\kappa S^3, 
\end{equation}
where we impose $Z_3$ symmetry to forbid liner and quadratic terms of $S$
and $H_{d, u}$. 
To eliminate the domain wall formed in the early Universe \footnote{See for example \cite{Hattori:2015xla} and its references}, $Z_3$ symmetry must be broken by some higher dimensional operators.
We also have the Yukawa couplings of quarks and leptons as in the MSSM.
In the following, we will use the same notation for the supermultiplet and its scalar component.
After the scalar component of $S$ develops a vacuum expectation value (VEV), the Higgs fields obtain the effective $\mu$ term, $\mu= \lambda\langle S \rangle$. The size of $\langle S \rangle$ is controlled by the soft SUSY breaking terms.
Thus as is well known, $\mu$ problem in the MSSM is solved in the NMSSM \cite{Kim:1983dt}. The $B$ term of the MSSM is also replaced with $A_\lambda$ and its size is the order of $M_0$ in the mirage mediation. Then the first ($B\mu$) problem of the mirage mediation is solved \cite{Choi:2005hd}.
In the NMSSM, we have a new quartic coupling, $\lambda^2 |H_d H_u|^2$ in the Higgs potential. This helps to raise the Higgs boson mass if $\tan\beta$ is small.
Also the mixing between the singlet and doublet Higgs fields pushes the mass of the SM-like Higgs boson if the singlet is lighter than the doublet.
These features are useful to solve the second problem in the mirage mediation.

We apply the TeV scale mirage mediation to the NMSSM \cite{Kobayashi:2012ee, Hagimoto:2015tua}.
We consider a model with the following modular weights for fields in the NMSSM,
\begin{eqnarray}
&& c_{H_u} = 0, \qquad c_{H_d} = 1, \qquad c_S = 0,\\
&& c_i = \frac{1}{2}, \qquad (i= Q, \overline{U}, \overline{D}, L, \overline{E}).
\label{eq:boundary-condition}
\end{eqnarray}
We assume the same modular weights for three generations of quarks and leptons. For the top Yukawa coupling and $\lambda$, the sum rule is satisfied, while it is violated for $\kappa$ and the bottom Yukawa coupling.
Then, we are only interested in the parameter region with small $\kappa$ and small/moderate $\tan\beta$.
In any case, we need small $\kappa$ to evade the Landau pole if we want to have large $\lambda$ for raising the Higgs mass.  The large $\tan\beta$ region is also excluded by the LEP chargino mass bound. Thus their corrections to the mirage unification are actually small.

We have chosen the above modular weights because the fine-tuning of the (effective) $\mu$ term is considerably ameliorated \cite{Choi:2006xb}. 
In the context of the natural SUSY \cite{natural-SUSY}, $|m_{H_u}^2|\sim \mu^2 \sim m_Z^2$ is often quoted as a spectrum to solve the fine-tuning problem in (\ref{eq:fine-tuning}) and predicts a light higgsino.
However, even if we have $|m_{H_u}^2|\sim m_Z^2$ with the heavy SUSY spectrum for some reason, we need another mechanism to fine-tune $\mu \sim m_Z$ because its origin is somehow related with the SUSY breaking and thus belongs to the heavy spectrum.
The doublet Higgs mass matrix of the model at the origin of the potential is given by,
\begin{equation}
 ({\cal M}_H^2)_{ij} =
\left(  
\begin{array}{cc}
M_0^2 + \mu^2 & M_0 \mu \\
M_0 \mu & \mu^2
\end{array}
\right),
\end{equation}
where $i,j =(H_d, H_u)$.
The trace of the mass matrix is given by ${\rm Tr}({\cal M}_H^2) = M_0^2+\mu^2$ while $M_0$ cancels in the determinant as ${\rm Det}({\cal M}^2_H) = \mu^4$. 
Thus the mass of the light doublet, $h$ is suppressed as $M_h \approx \mu (\mu/M_0)$, while that of the heavy doublet, $H$ is $M_H \approx M_0$.
This corresponds to the cancellation of the second and the third terms in the formula,
\begin{equation}
 \frac{m_Z^2}{2}
 \simeq -m^2_{H_u} -\mu^2 +\frac{m_{H_d}^2}{\tan^2\beta}. 
\label{eq:mZ}
\end{equation}
Thus the electroweak scale $m_Z$ is insensitive to $\mu$ in the model and
 mostly determined by $|m_{H_u}^2|$.
$\mu$ can be as large as $\sqrt{m_Z M_0}$ without deteriorating the fine-tuning.
The sum rule $A_\lambda= M_0$ in the NMSSM and $m_{H_d}^2=M_0^2$ are essential for this cancellation. Note that a similar mechanism works in the MSSM, however, the $B$ term is a remnant of the cancellation and vulnerable to uncontrollable corrections.

In the NMSSM, mixing with the singlet may destroy the above nice feature.
In addition it also causes phenomenological problems.
If the lightest CP even Higgs boson is doublet (SM) like, the mixing with
 the singlet reduces the Higgs boson mass and makes it
 difficult to achieve 125 GeV. 
If the lightest CP even Higgs boson is singlet like, the mixing with the doublet tightens the constraint from the LEP Higgs boson search because it couples with the $Z$ boson via the mixing.
However, we find that, in the model with $\kappa<<1$ and $m_{S, H_u}^2<< m_{H_d}^2$ like we are considering, this mixing is suppressed due to approximate scale symmetries \cite{Hagimoto:2015tua}.
In the limit, $\kappa=0$ and $m_S^2=0$, the model has the following approximate scale symmetry,
\begin{eqnarray}
H_u(x) &=& e^{2\phi} H_u^\prime(e^\phi x), \\
H_d(x) &=& e^{2\phi} H_d^\prime(e^\phi x), \\
S(x) &=& S^\prime(e^\phi x),  
\end{eqnarray} 
which is explicitly broken by the K\"ahler potential $S^\dag S$, the D-term potential and all the kinetic terms.
The VEVs of $H_{d,u}$ breaks the symmetry and the light doublet is the corresponding Nambu-Goldstone (NG) boson which is inherently a mass eigen state. Thus the mass and the mixing with the singlet must pick up the explicit breaking.
While in the $\kappa=0$ and $m_{H_u}^2=0$ limit, the model has another approximate symmetry in which $S$ and $H_u$ are exchanged in the above formulas. 
The NG boson is a mixture of $S$ and $H_d$ and its mass and the mixing with $H_u$ also need to pick up the explicit breaking.
Thus the mixing between the light doublet and the singlet must pick up the explicit breaking terms of the both symmetries or a term breaking both of them because it vanishes if either of the symmetries survives.
In addition, the mixing disappears if $H_d$ decouples and it must be suppressed by $m_{H_d}^2$ (Thus, only $\kappa$ or the gauge coupling is not enough to generate the mixing). Therefore the doublet-singlet mixing is expected to be highly suppressed at classical level in the model.

\section{The SM-like Higgs boson mass}

In the following we show the results of our numerical calculations.
We solve the one-loop renormalization group equations from the unification scale to the SUSY scale, assuming the boundary condition (\ref{eq:boundary-condition}) in addition with the contribution from the anomaly mediation. 
We adopt these solutions for the SUSY spectrum and the large parameters, however, discard them for the small parameters, $m_{H_u}^2$, $m_{S}^2$, $\mu$ and $A_\kappa$. We rather deal them as free parameters at the SUSY scale, taking into account the ambiguities stemming from the threshold corrections and two-loop renormalization group running. Then we solve the minimum of the one-loop effective Higgs potential in terms of them so that we obtain the observed value of $m_Z$. We accept the solutions if they are within the range of the ambiguities. We fix $A_\kappa$ to a reference value $-100$ GeV because there are not enough equations to determine it. We use the \texttt{NMSSMTools} package for calculating the physical mass spectrum \cite{Ellwanger:2005dv}.
We have three CP-even, two CP-odd and one charged Higgs bosons.
The heaviest CP-even, CP-odd and charged Higgs bosons correspond to the heavy doublet and have almost degenerate masses $\approx M_0$.
We have two CP-even and one CP-odd Higgs bosons at the electroweak scale.
We chose the default option of the \texttt{NMSSMTools} to estimate their mass spectrum.

\begin{figure}[htbp]
\begin{center}
\begin{tabular}{l @{\hspace{10mm}} r}
\includegraphics[height=50mm]{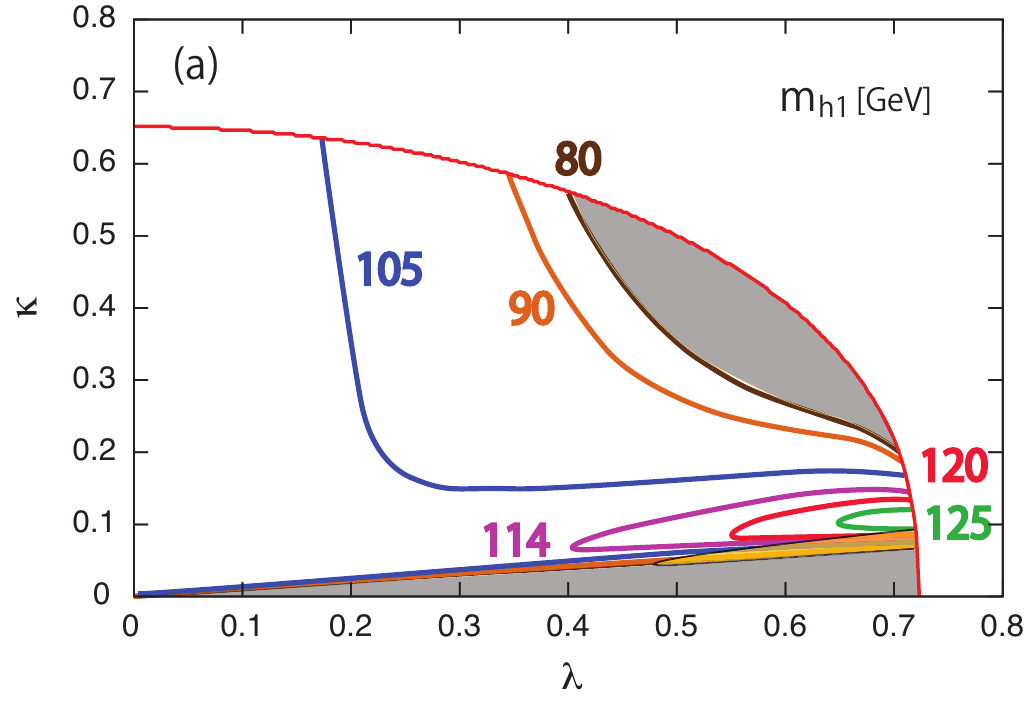} &
\includegraphics[height=50mm]{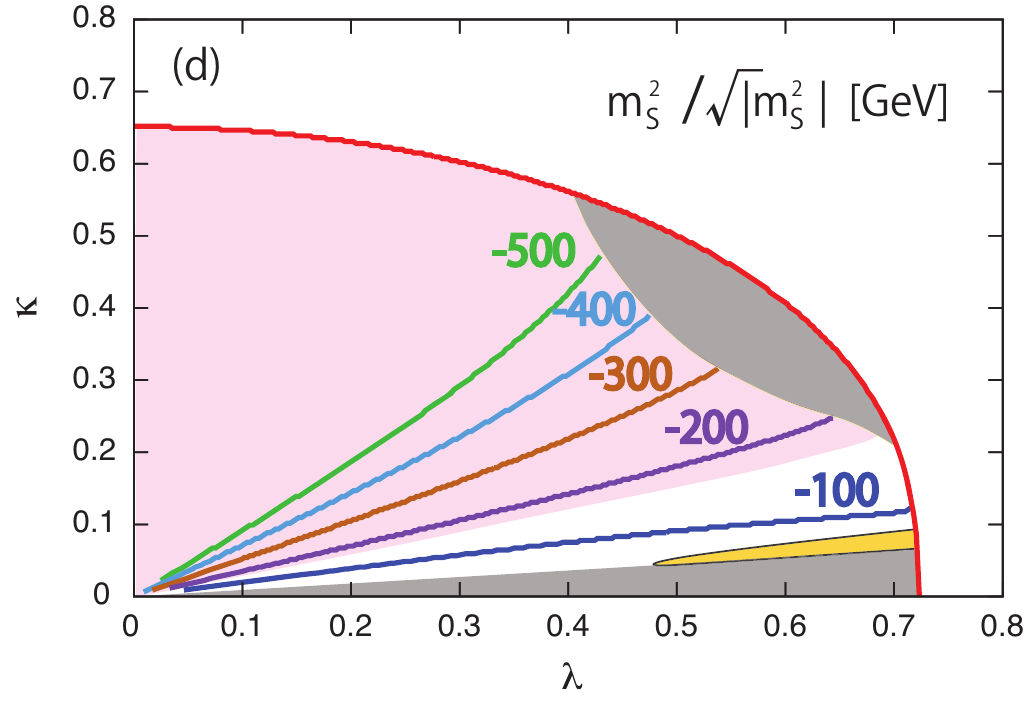} \\
\rule{0cm}{10mm} & \\
\includegraphics[height=50mm]{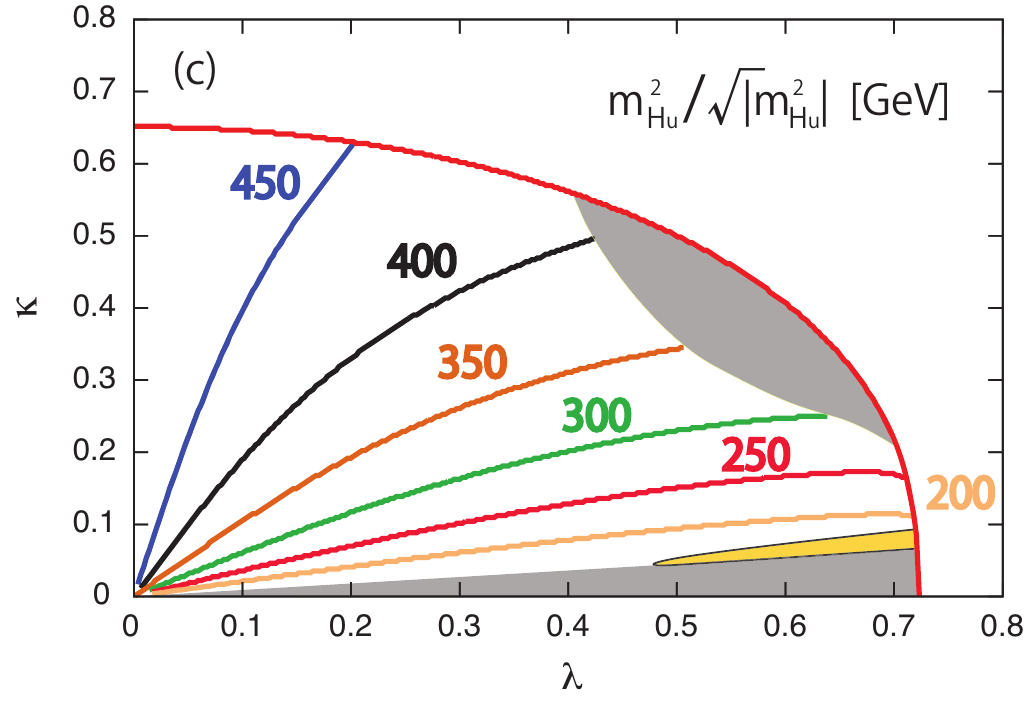} &
\includegraphics[height=50mm]{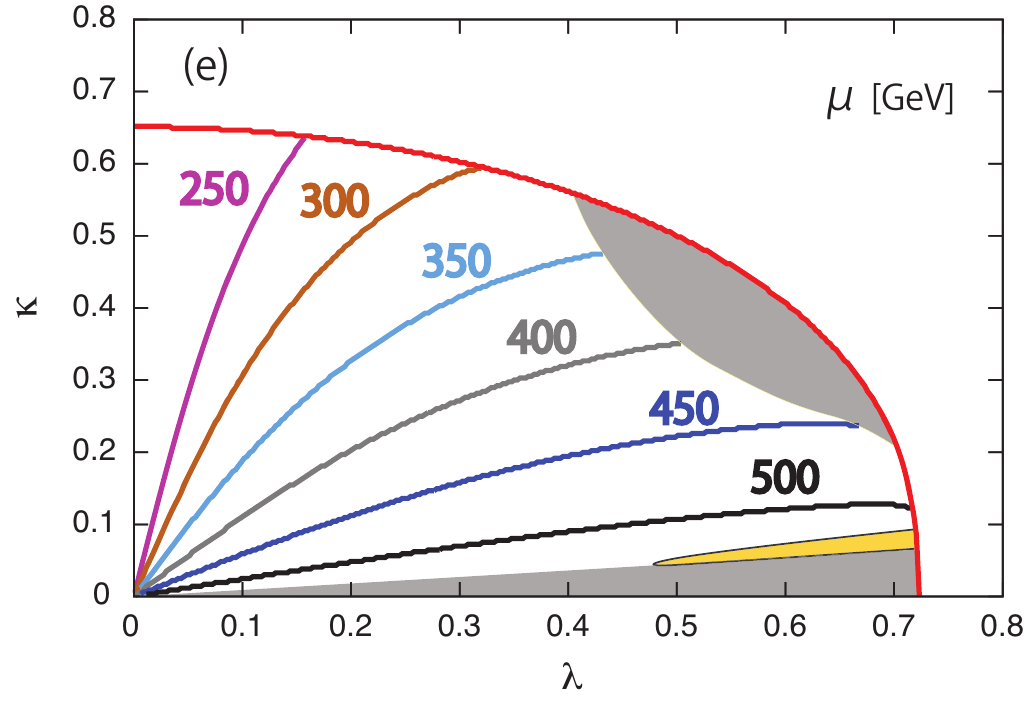} \\ 
\end{tabular}
\caption{
The SM-like Higgs boson mass and the small parameters for 
$\tan\beta=3$, $M_0=1500$ GeV and $A_\kappa = -100$ GeV.
\label{fig:contour-tanb3}}
\end{center}
\end{figure}

The figure \ref{fig:contour-tanb3} shows the contour plots of the SM-like Higgs boson mass and the small parameters for $\tan\beta=3$, $M_0=1500$ GeV and $A_\kappa = -100$ GeV. In this parameter region, the lightest CP-even Higgs boson is the SM like.
In the plots, the red curve indicate that, outside of it, $\lambda$ or $\kappa$ hits the Landau-pole below the unification scale.
The gray region on the bottom of the plots is excluded because the Higgs boson is tachyonic. 
In another gray region around the red curve, an iterative method to solve the Higgs mass fails. We do not take care of this since the region is already excluded by the LEP Higgs boson search.   
In the yellow region the SM like vacuum is metastable \cite{Kanehata:2011ei}.
The upper-left plot shows the Higgs boson mass reaches 125 GeV around $\lambda \simeq 0.7$ and $\kappa \simeq 0.1$.
In the upper-right plot, the white region is the prediction of $m_S^2$ in the TeV scale mirage mediation with the ambiguities coming from the threshold corrections and the 2-loop running. It overlaps with the region predicting the 125 GeV Higgs boson. In the lower-left plot, we confirm that $m_{H_u}^2$ is also within the model ambiguity, $m_{H_u}^2 \lesssim M_0^2/8\pi^2$ around this region.
In the lower-right plot, we see that $\mu \approx M_0/\tan\beta \approx 500$ GeV 
around the region where $m_{h_1}=125$ GeV and is favored by the model.

\begin{figure}[htbp]
\begin{center}
\begin{tabular}{l @{\hspace{10mm}} r}
\includegraphics[height=50mm]{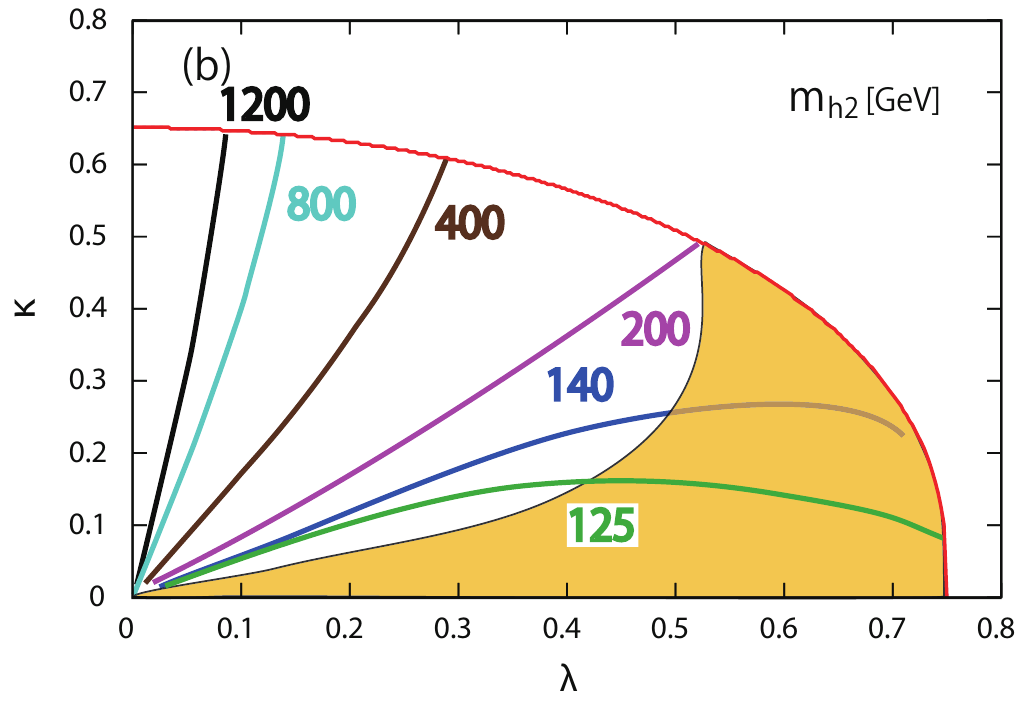} &
\includegraphics[height=50mm]{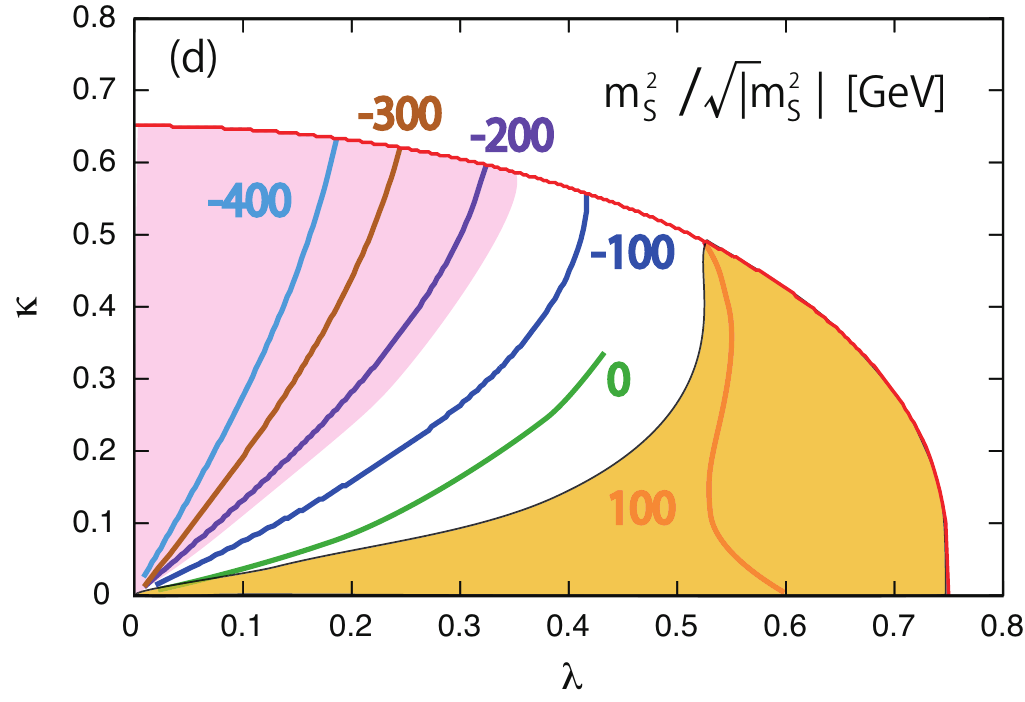} \\
\rule{0cm}{10mm} & \\
\includegraphics[height=50mm]{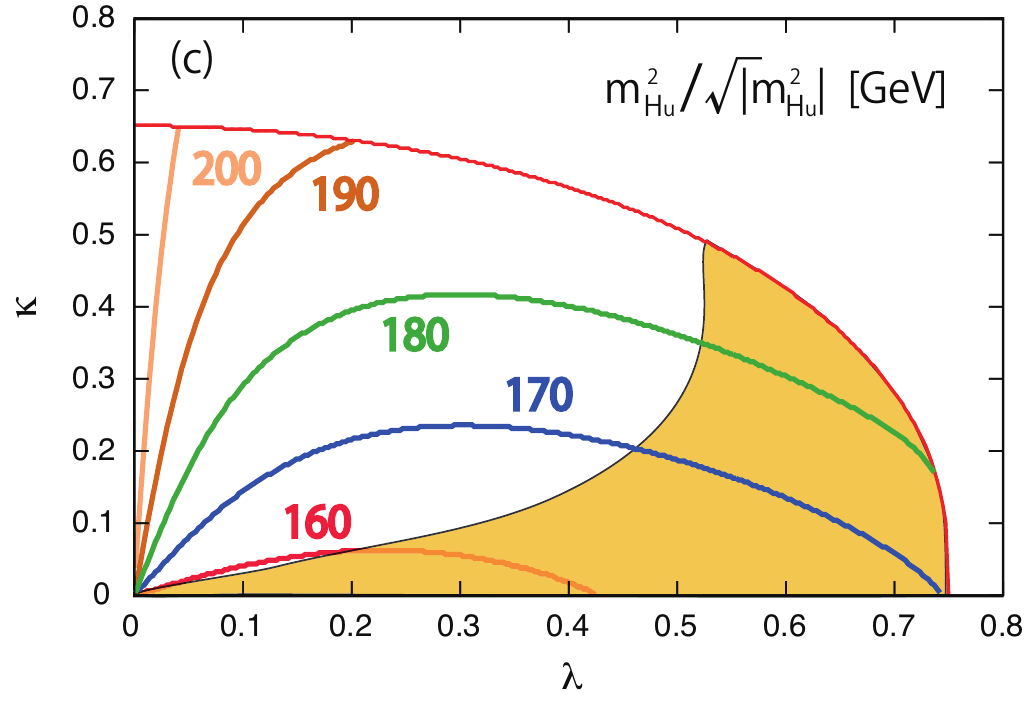} &
\includegraphics[height=50mm]{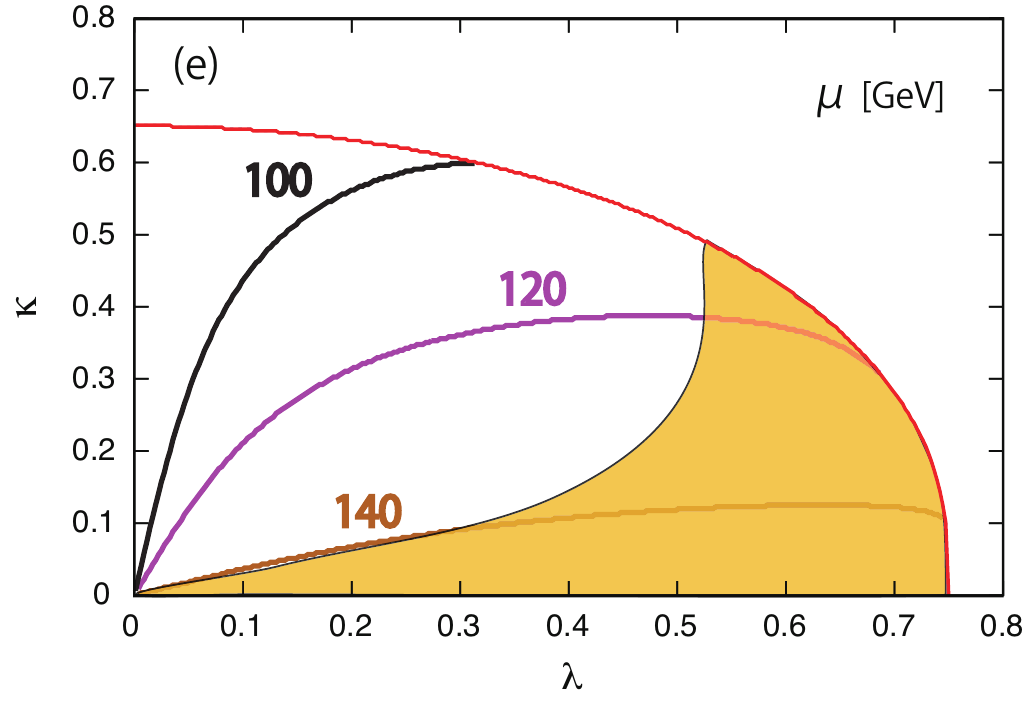} \\ 
\end{tabular}
\caption{
The SM-like Higgs boson mass and the small parameters for 
$\tan\beta=10$, $M_0=1500$ GeV and $A_\kappa = -100$ GeV.
\label{fig:contour-tanb10}
}
\end{center}
\end{figure}

The figure \ref{fig:contour-tanb10} shows the similar plots for $\tan\beta=10$.
The other parameters are the same as those in the figure \ref{fig:contour-tanb3}. In this parameter region, the second lightest CP-even Higgs boson is the SM like. Again, we confirm that $m_S^2$ and $m_{H_u}^2$ are within the prediction of the TeV scale mirage mediation in the region where $m_{h_2}= 125$ GeV.
The corresponding $\kappa$ is small enough to ensure the mirage unification
 and $\mu \approx M_0/\tan\beta \approx 150$ GeV is also satisfied, displaying the $\mu$ cancellation in (\ref{eq:mZ}) is working there.

\section{Fine-tuning of the electroweak symmetry breaking}

\begin{figure}[htb]
\begin{center}
\begin{tabular}{l @{\hspace{10mm}} r}
\includegraphics[height=45mm]{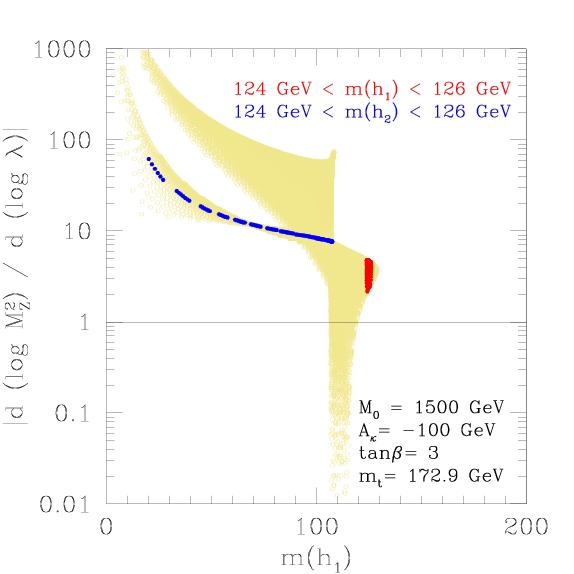} &
\includegraphics[height=45mm]{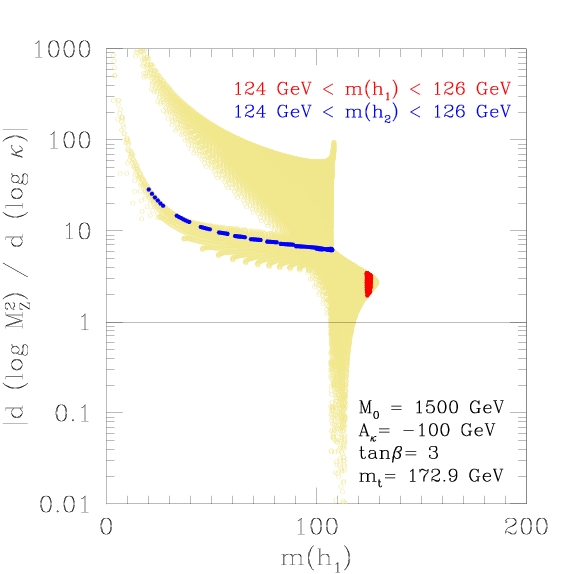} \\ 
\end{tabular}
\begin{tabular}{l @{\hspace{10mm}} c @{\hspace{10mm}} r}
\includegraphics[height=45mm]{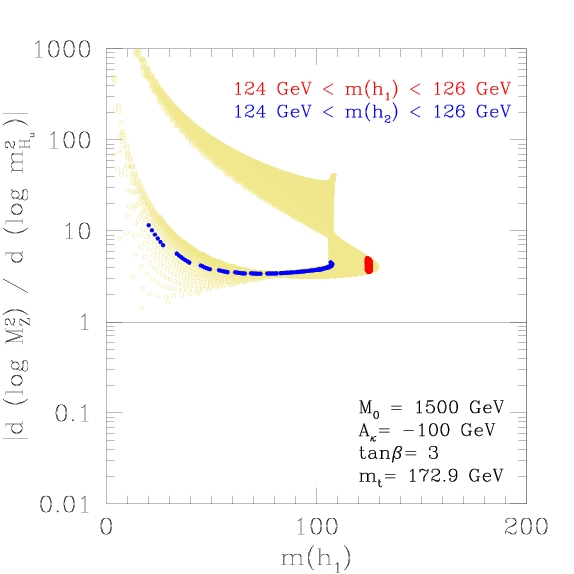} &
\includegraphics[height=45mm]{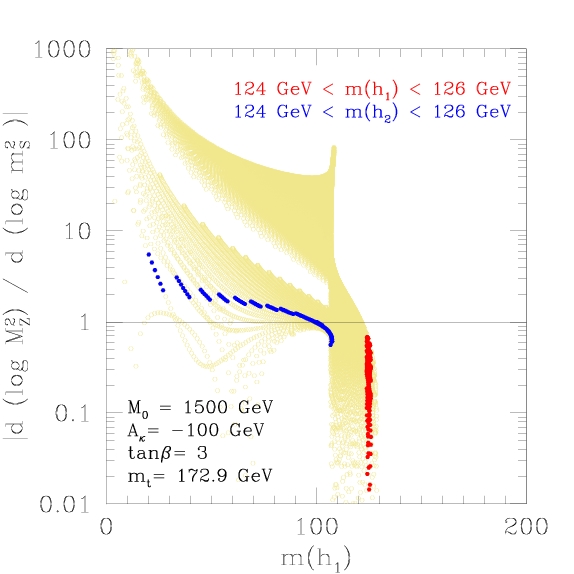} & 
\includegraphics[height=45mm]{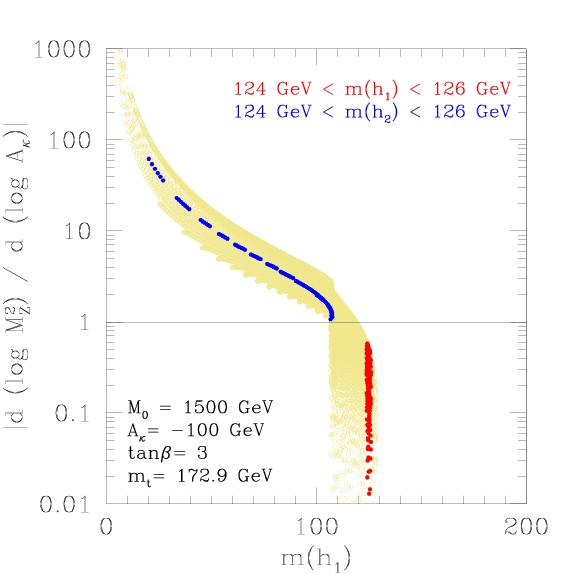} \\ 
\end{tabular}
\caption{
The fine-tuning measures, $\Delta^{m_Z^2}_X$ for $\tan\beta=3$, $M_0=1500$ GeV and $A_\kappa=-100$ GeV.
\label{fig:finetuning-mz-tanb3}
}
\end{center}
\end{figure}

\begin{figure}[htb]
\begin{center}
\begin{tabular}{l @{\hspace{10mm}} r}
\includegraphics[height=45mm]{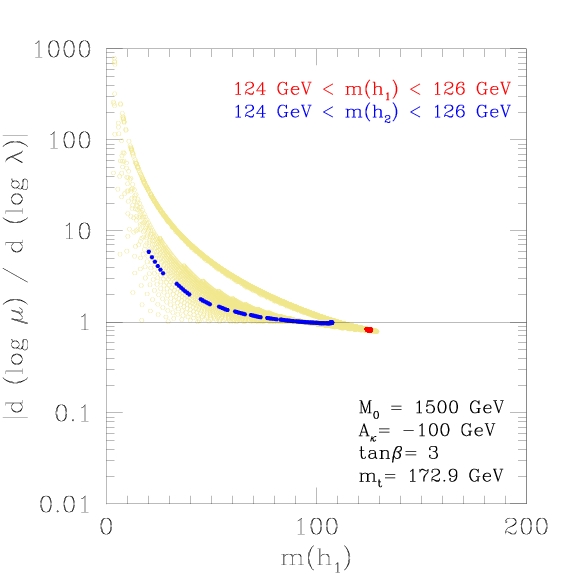} &
\includegraphics[height=45mm]{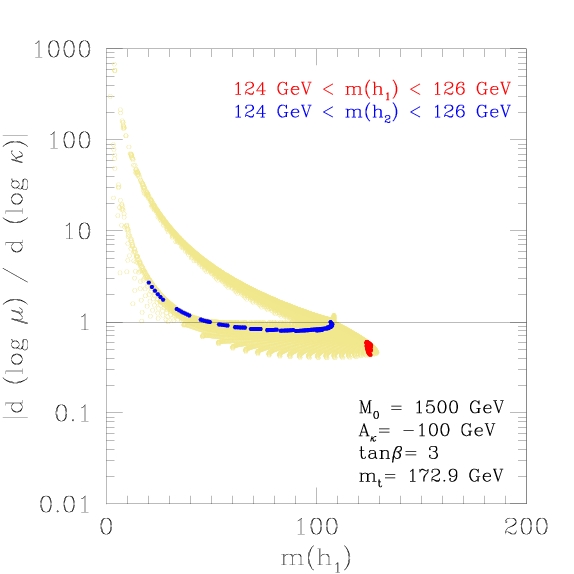} \\ 
\end{tabular}
\begin{tabular}{l @{\hspace{10mm}} c @{\hspace{10mm}} r}
\includegraphics[height=45mm]{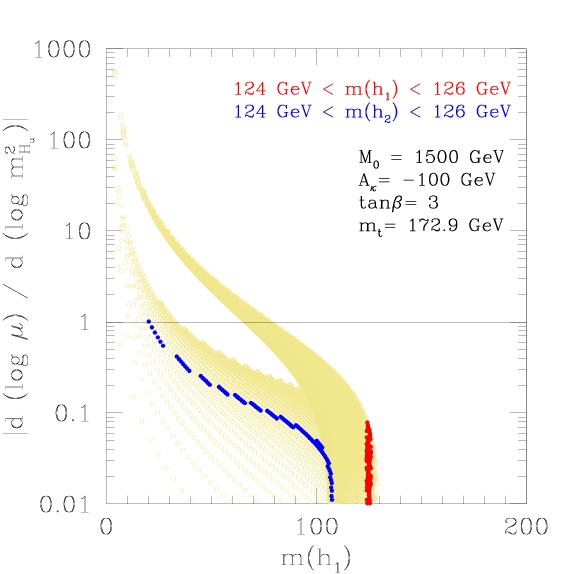} &
\includegraphics[height=45mm]{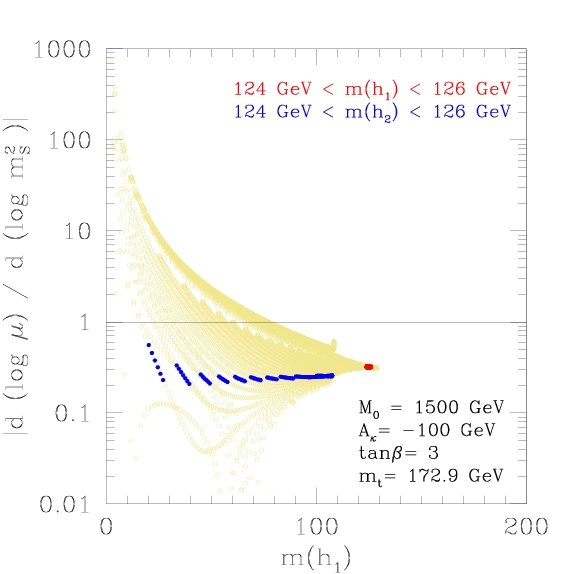} & 
\includegraphics[height=45mm]{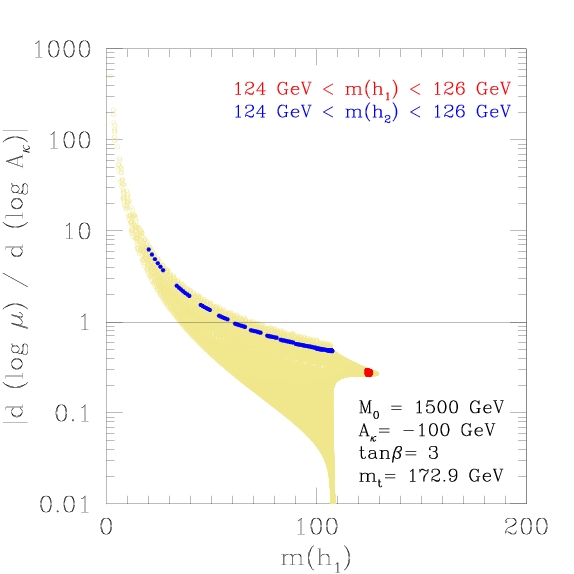} \\ 
\end{tabular}
\caption{
The fine-tuning measures, $\Delta^{\mu}_X$ for $\tan\beta=3$, $M_0=1500$ GeV and $A_\kappa=-100$ GeV.
\label{fig:finetuning-mu-tanb3}}
\end{center}
\end{figure}

Now we confirmed that the observed Higgs boson mass is realized with the TeV scale mirage mediation in the NMSSM, next we estimate the degree of fine-tuning of the electroweak symmetry breaking in this model.
First we define the fine-tuning measures.
We define them as following,
\begin{equation}
 \Delta^y_x = \frac{\partial \ln(y)}{\partial \ln(x)},
\end{equation}
where x denotes the input parameter and y denotes the output observable of the model \cite{FT, FTLEP}. Natural choice for $y$ is the Higgs VEVs: $\langle H_u\rangle$, $\langle H_d \rangle$ and $S$. Instead, we choose $m_Z^2$, $\tan\beta$, $\mu$ extending the standard choice, $m_Z^2$ in literature. We stress that all the VEVs must be considered once we extend the Higgs sector. As for the input parameter $x$, we take the small parameters.
Note that the large parameters like $m_{H_d}^2$ and $A_\lambda$ are fixed by the UV physics and not free parameters. The worst measure, $\Delta = {\rm max} (\Delta^y_x)$ might be considered as the fine-tuning measure of the model.
The measures can be estimated using the following master formula \cite{Hagimoto:2015tua},
\begin{equation}
 \frac{\partial v_i}{\ln x_a}=-\frac{1}{\sqrt{2}} \sum_k ({\cal M}_S^2)^{-1}_{ik} \frac{\partial^2 V(\phi)}{\partial \phi_k \partial \ln x_a},
\end{equation}
where ${\cal M}_S^2$ denotes the CP-even Higgs boson mass matrix and $V$ denotes the Higgs potential.
The physical meaning of this equation is obvious that the fine-tuning gets worse if a certain Higgs boson becomes significantly lighter than the parameters in the Higgs potential. We use the loop corrected ${\cal M}_S^2$ for the calculation of the measures instead of the tree-level estimation in the \texttt{NMSSMTools}.

Figure \ref{fig:finetuning-mz-tanb3} shows $\Delta^{m_Z^2}_x$ for $\tan\beta=3$, $M_0=1500$ GeV and $A_\kappa=-100$ GeV. 
In the plots, the red points satisfy $124\, {\rm GeV} < m_{h_1} < 126\, {\rm GeV}$ and also the requirement of the TeV scale mirage mediation, as we saw in the previous section. 
On the other hand, the blue points indicate $124\, {\rm GeV} < m_{h_2} < 126\, {\rm GeV}$, although they are already excluded by the LEP Higgs boson search \cite{Schael:2006cr}. In both cases, the fine-tuning measures are better than 10 unless $m_{h_1}$ is much lighter than the electroweak scale. This low fine-tuning is not a trivial result because the khaki region which is not necessarily favored by the TeV scale mirage mediation (but not far from it) reaches 100 even with a moderate value of $m_{h_1}$.
Figure \ref{fig:finetuning-mu-tanb3} shows the similar plots for $\Delta^{\mu}_x$. All of them are below the unity if $m_{h_1}$ is around the electroweak scale.
The remaining measures, $\Delta^{\tan\beta}_X$ are not independent of $\Delta^{\mu}_X$ in our model because $\mu \approx M_0/\tan\beta$ holds very well, which ensures the $\mu$ cancellation mechanism.

\begin{figure}[htb]
\begin{center}
\begin{tabular}{l @{\hspace{10mm}} r}
\includegraphics[height=60mm]{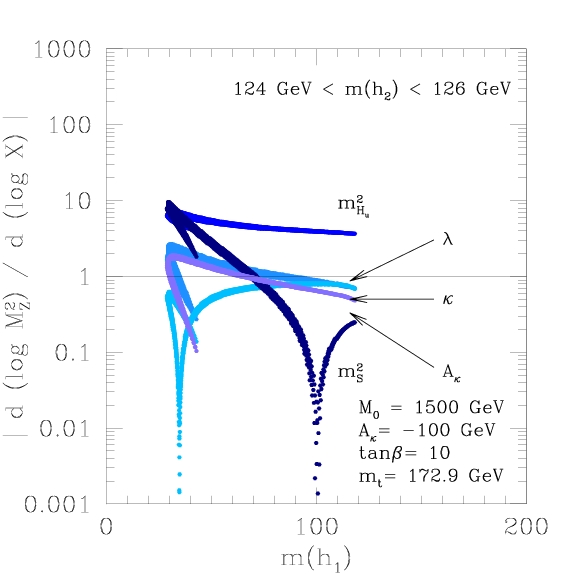} &
\includegraphics[height=60mm]{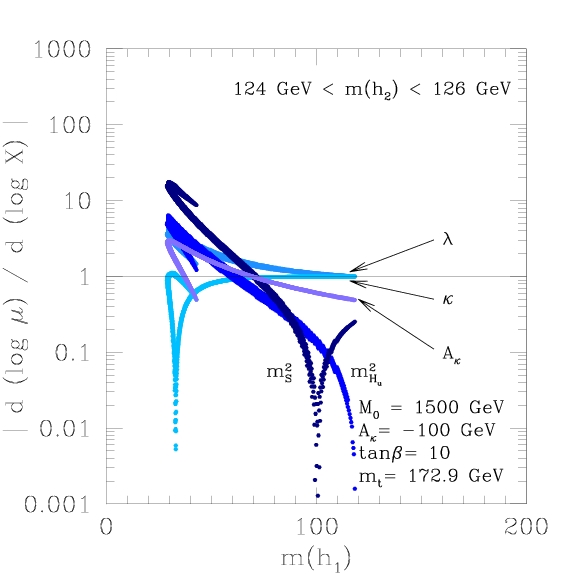} \\ 
\end{tabular}
\caption{
The fine-tuning measures for $\tan\beta=10$, $M_0=1500$ GeV and $A_\kappa=-100$ GeV.
\label{fig:finetuning-tanb10} 
}
\end{center}
\end{figure}

Figure \ref{fig:finetuning-tanb10} shows the fine-tuning measures for $\tan\beta=10$. The other parameters are same as in the $\tan\beta=3$ case. Here, only the points satisfying $124\, {\rm GeV} < m_{h_2} < 126\, {\rm GeV}$ are plotted. For this $\tan\beta$, no parameter region realizes $124\, {\rm GeV} < m_{h_1} < 126\, {\rm GeV}$. Almost all the points are below 10 unless $m_{h_1}$ is too light.
Note that the fine-tuning is most relaxed around $m_{h_1}\simeq 100$ GeV
 which is exactly the region surviving from the LEP Higgs boson search as we will see later \cite{Schael:2006cr}.

\begin{figure}[htb]
\begin{center}
\begin{tabular}{l @{\hspace{10mm}} r}
\includegraphics[height=60mm]{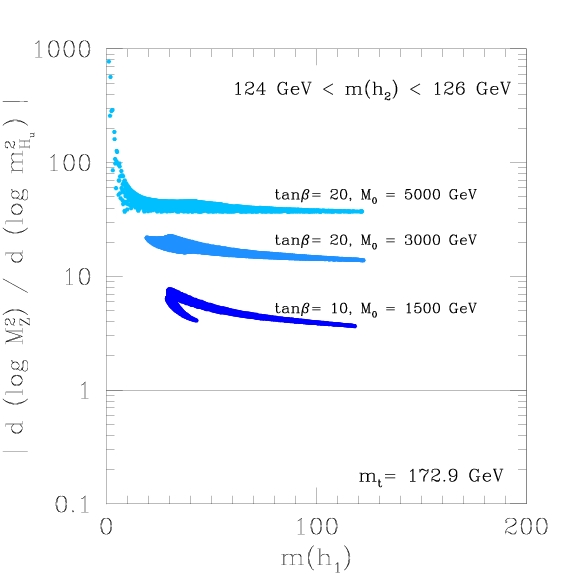} &
\includegraphics[height=60mm]{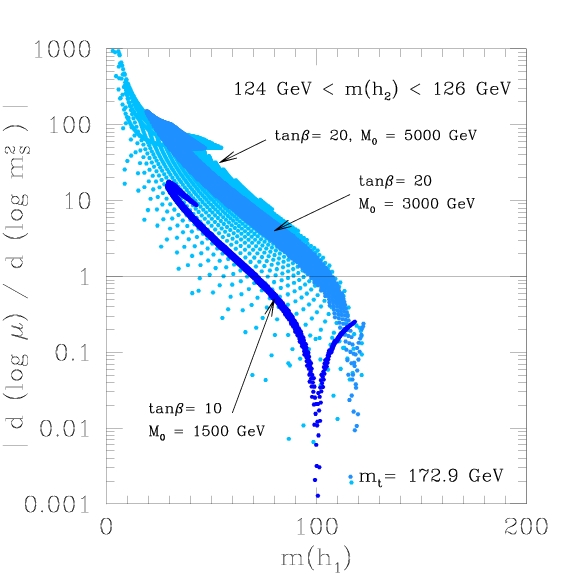} 
\end{tabular}
\end{center}
\caption{The fine-tuning measures $\Delta^{m_Z^2}_{m_{H_u}^2}$ (left) and $\Delta^\mu_{m_S^2}$ (right) for $M_0=3$ TeV and $M_0=5$ TeV ($\tan\beta=20$) compared with $M_0=1.5$ TeV ($\tan\beta=10$) case. $A_\kappa$ is fixed at $A_\kappa=-100$ GeV. \label{fig:heavy-case}}
\end{figure}

With the above numerical analysis, we conclude that our model realizes the electroweak symmetry breaking with better than 10\% tuning for 1.5 TeV gluino and 1 TeV stop masses. 
For completeness, we raise the SUSY scale, $M_0$ up to $5$ TeV in figure \ref{fig:heavy-case}. It is surprising that the worst measure $\Delta^{m_Z^2}_{m_{H_u}^2}$ is around 40 at $M_0=5$ TeV which is acceptable in the standard of conventional 1 TeV models. It is also noted that $\Delta^{\mu}_{m_S^2}$ is again most relaxed around $m_{h_1}\simeq 100$ GeV where the constraint from the LEP Higgs boson search is weak due to anomalous events \cite{Belanger:2012tt,Drees:2005jg,Dermisek:2005gg}.

\section{Precision Higgs measurement}

In this section, we discuss the coupling constants of the Higgs bosons in our model and the prospects for their precision measurement. 
The effective coupling constants of the CP-even Higgs bosons are defined as \cite{Carmi:2012in},
\begin{eqnarray}
{\cal L} &=& \sum_{i=1}^3
\left[
 C_V^i \frac{\sqrt{2} m_W^2}{v} h_i W_\mu^+ W^{-\mu}
+C_V^i \frac{m_Z^2}{\sqrt{2}v} h_i Z_\mu Z^\mu
-\sum_f C_f^i \frac{m_f}{\sqrt{2} v} h_i \overline{f} f \right. \nonumber\\
&& \left. +C_g^i \frac{\alpha_S}{12\sqrt{2}\pi v} h_i G_{\mu\nu}^a G^{a\,\mu\nu}
+C_\gamma^i \frac{\alpha}{\sqrt{2}\pi v} h_i A_{\mu\nu} A^{\mu\nu}
\right],
\end{eqnarray}
where we assume the custodial symmetry which relates the W and Z coupling constants. For the SM, they are given by $C_V^{SM}=C_f^{SM}=1$, $C_g^{SM}\approx 1.03$ and $C_\gamma^{SM} \approx -0.81$.
Deviations in the couplings of the SM-like Higgs boson or the mass and coupling of the new boson encode information of the new physics and now become important targets of the future lepton colliders after the LHC Run I excluded low mass colored new particles.

\begin{figure}[htb]
\begin{center}
\begin{tabular}{l @{\hspace{10mm}} r}
\includegraphics[height=60mm]{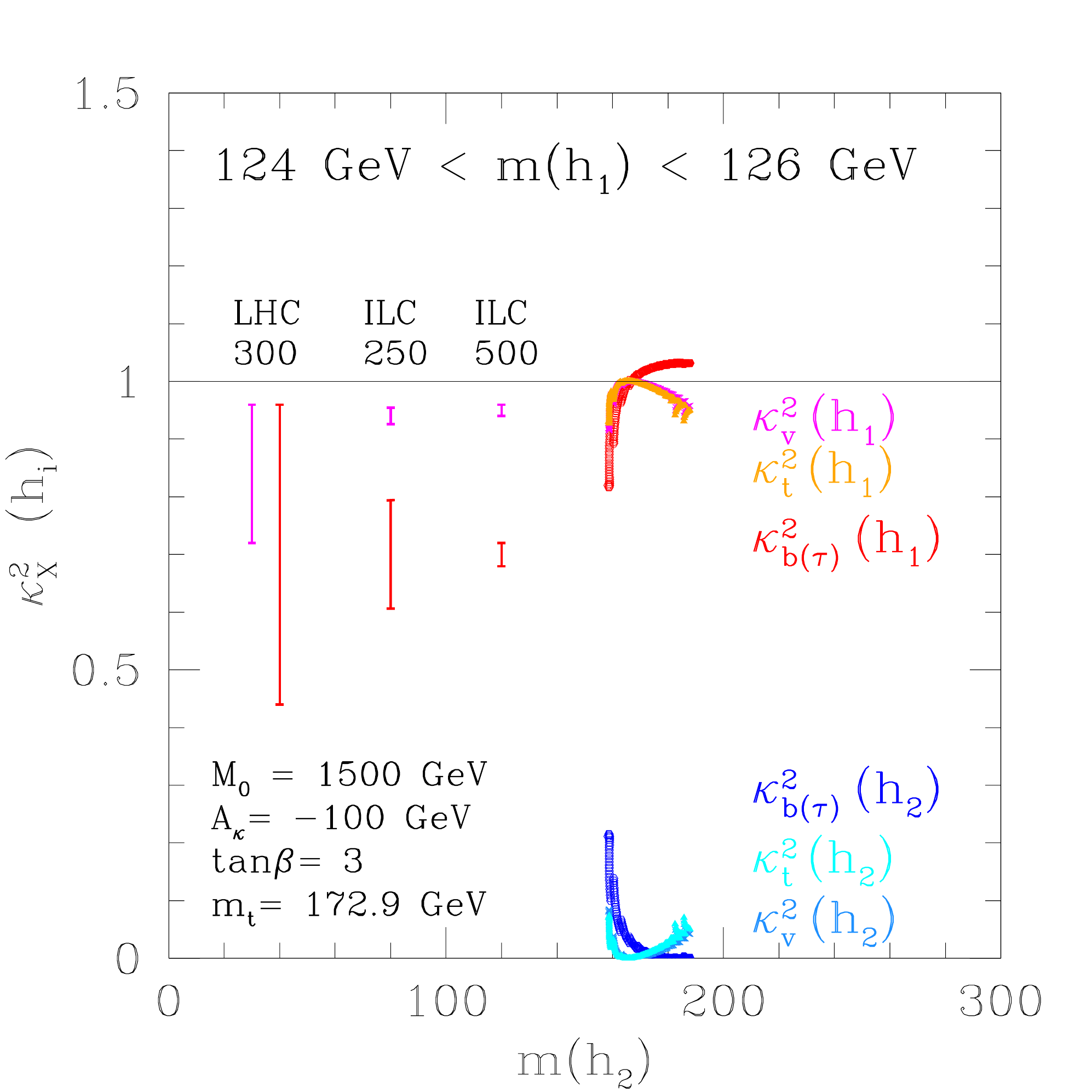} &
\includegraphics[height=60mm]{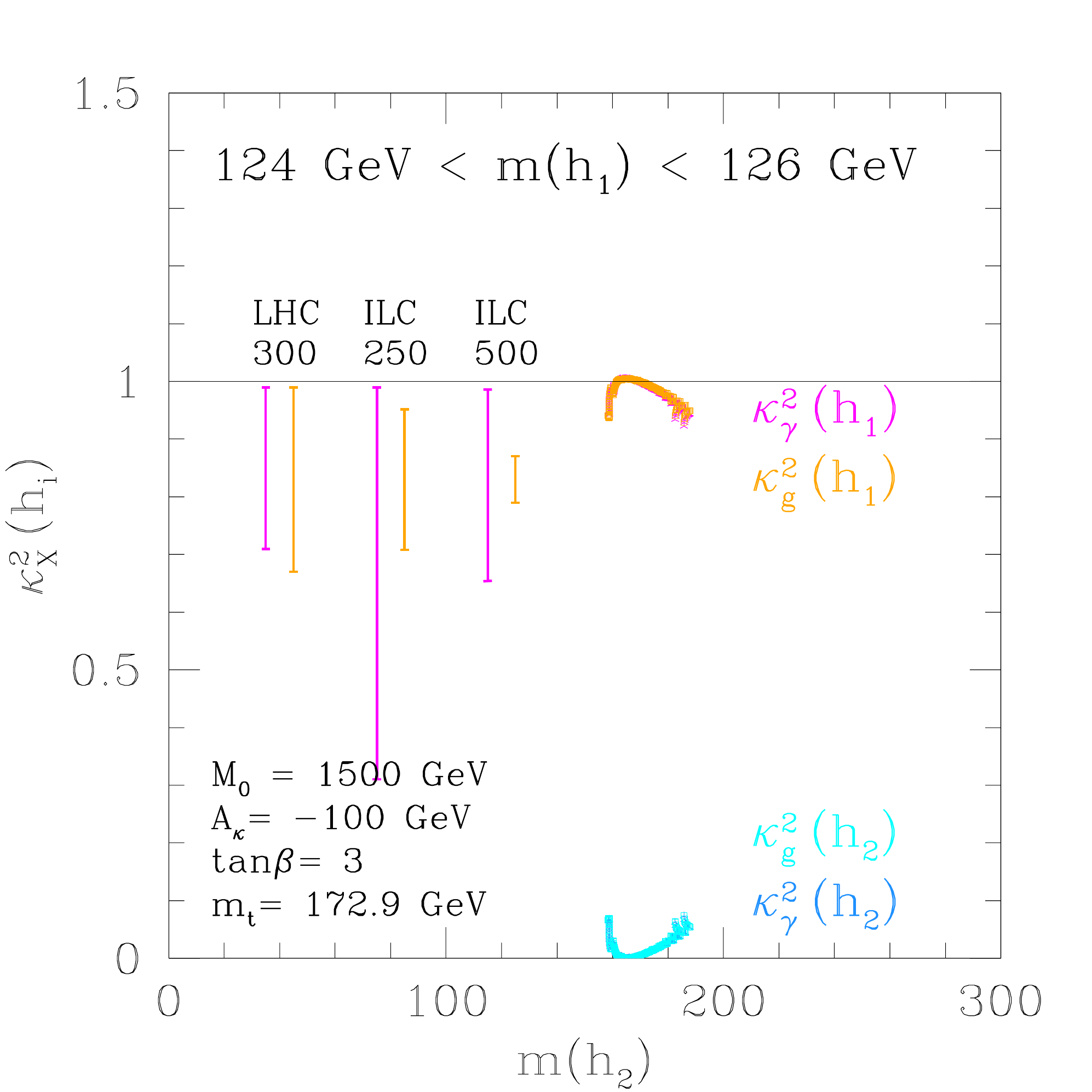} \\ 
\end{tabular}
\caption{\label{fig:cpevencouplings-tanb3} The scale factors of the CP-even Higgs coupling constants for $\tan\beta=3$, $M_0=1500$ GeV and $A_\kappa=-100$ GeV. The vertical lines show the expected precision of the SM-like Higgs coupling constants for LHC and ILC, $\Delta (\kappa_X)^2 \approx 2 \Delta \kappa_X$.}
\end{center}
\end{figure}

In the following numerical analysis, we use the scaling factor $\kappa_X^i$ instead of the coupling constant itself, which is defined as \cite{LHCHiggsCrossSectionWorkingGroup:2012nn, Heinemeyer:2013tqa},
\begin{equation}
 \kappa_X^i =\frac{C_X^i}{C_X^{SM}}.
\end{equation} 
The calculation is performed by the \texttt{NMSSMTools} package.
In figure \ref{fig:cpevencouplings-tanb3}, we plot the square of the scaling factors for $\tan\beta=3$, $M_0=1500$ GeV and $A_\kappa=-100$ GeV.
Here the lightest CP-even Higgs boson is the SM like.
With the vertical lines, we indicate the precision of their measurement at LHC (300 $fb^{-1}$) \cite{Dawson:2013bba}, ILC 250 GeV and ILC 500 GeV \cite{Asner:2013psa}. The positions of them are arbitrary.  Because of the little hierarchy, the loop corrections from the SUSY particles are negligible. Thus the deviations come from the mixing between the light doublet and singlet Higgs bosons. 
Consequently, a sum rule $(\kappa^1)^2+(\kappa^2)^2=1$ holds well except for the bottom quark or the tau lepton, for which the mixing with the heavy doublet is not negligible due to the $\tan\beta$ enhanced couplings.
According to the figure, the approximate scale symmetries suppress the deviation in the scaling factors of the SM-like Higgs boson below 5\% for the vector bosons, top quark (left panel) and loop induced couplings (right panel). ILC can resolve some of them but it is hard for LHC.
The deviations for the bottom quark and tau lepton are below 10\% and ILC will be able to investigate their detail. 
On the other hand, the couplings of the singlet-like Higgs boson are below 30\% of those for the SM Higgs boson for the vector bosons, top quark and loop induced couplings. 
The couplings for the bottom quark and tau lepton can be 50 \% of those for the SM Higgs boson.
LHC might be able to detect them. However ILC is most suitable to find the hidden new boson and examine its precise coupling strengths thanks to its background free nature.

\begin{figure}[htbp]
\begin{center}
\begin{tabular}{l @{\hspace{10mm}} r}
\includegraphics[height=60mm]{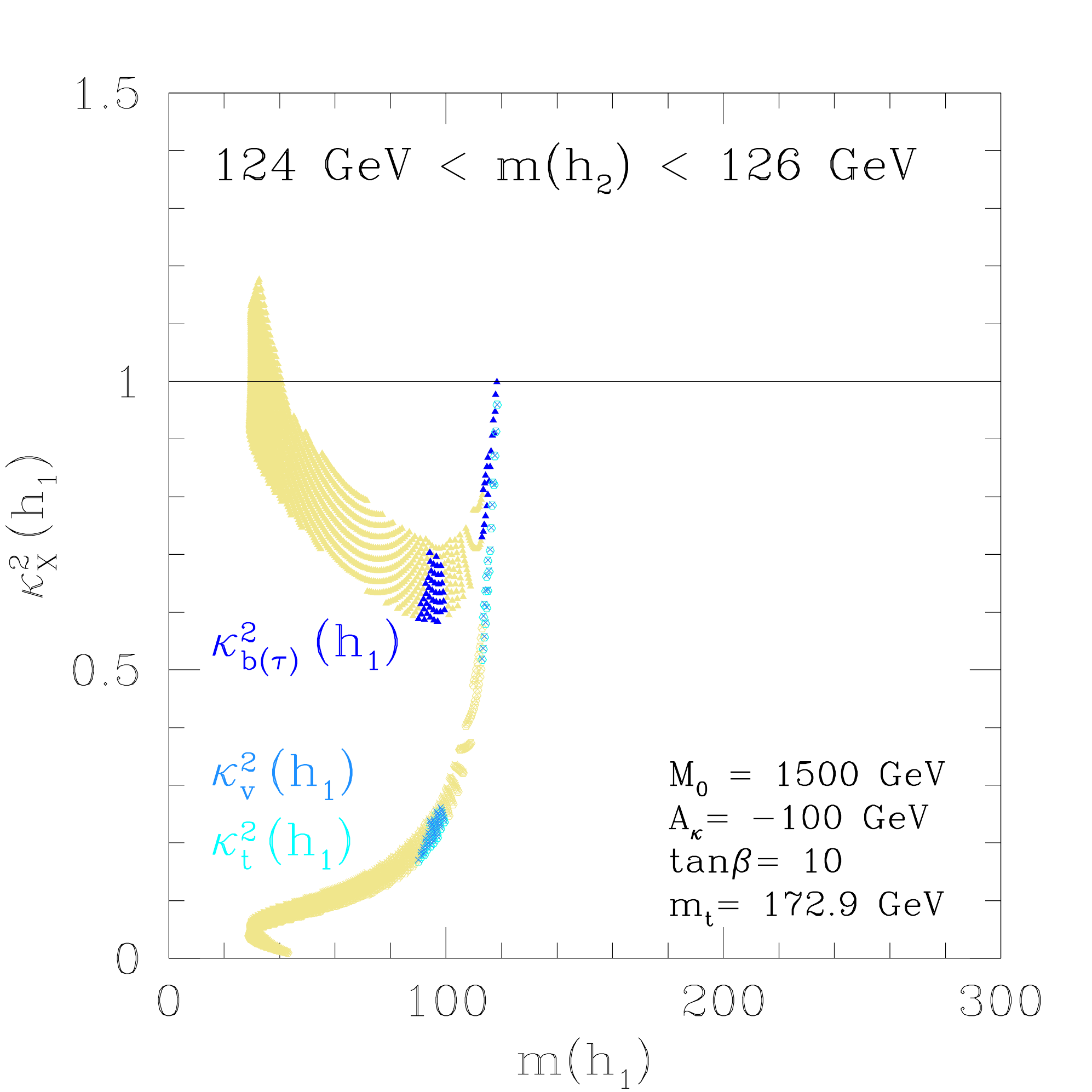} &
\includegraphics[height=60mm]{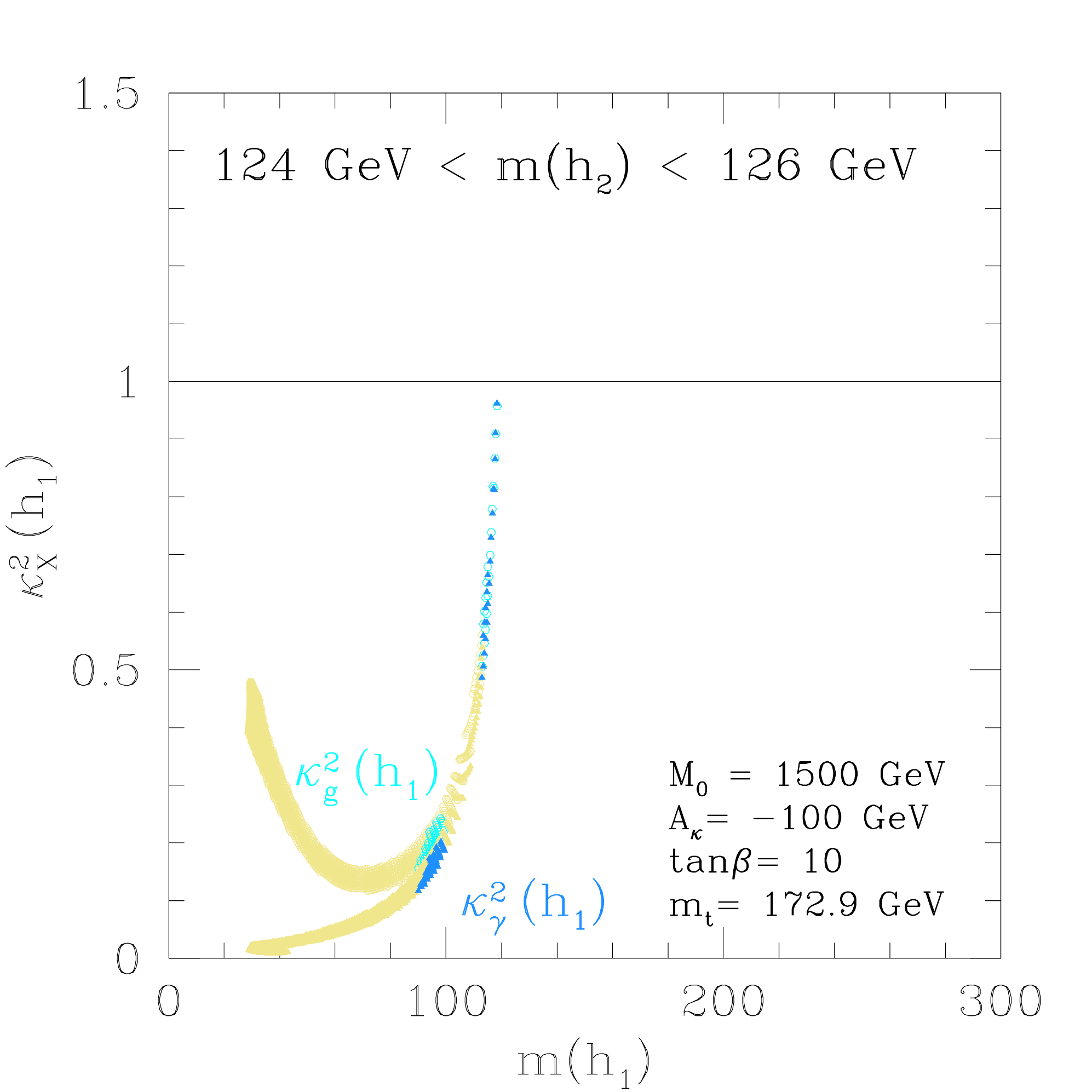} \\
\rule{0cm}{10mm} & \\
\includegraphics[height=60mm]{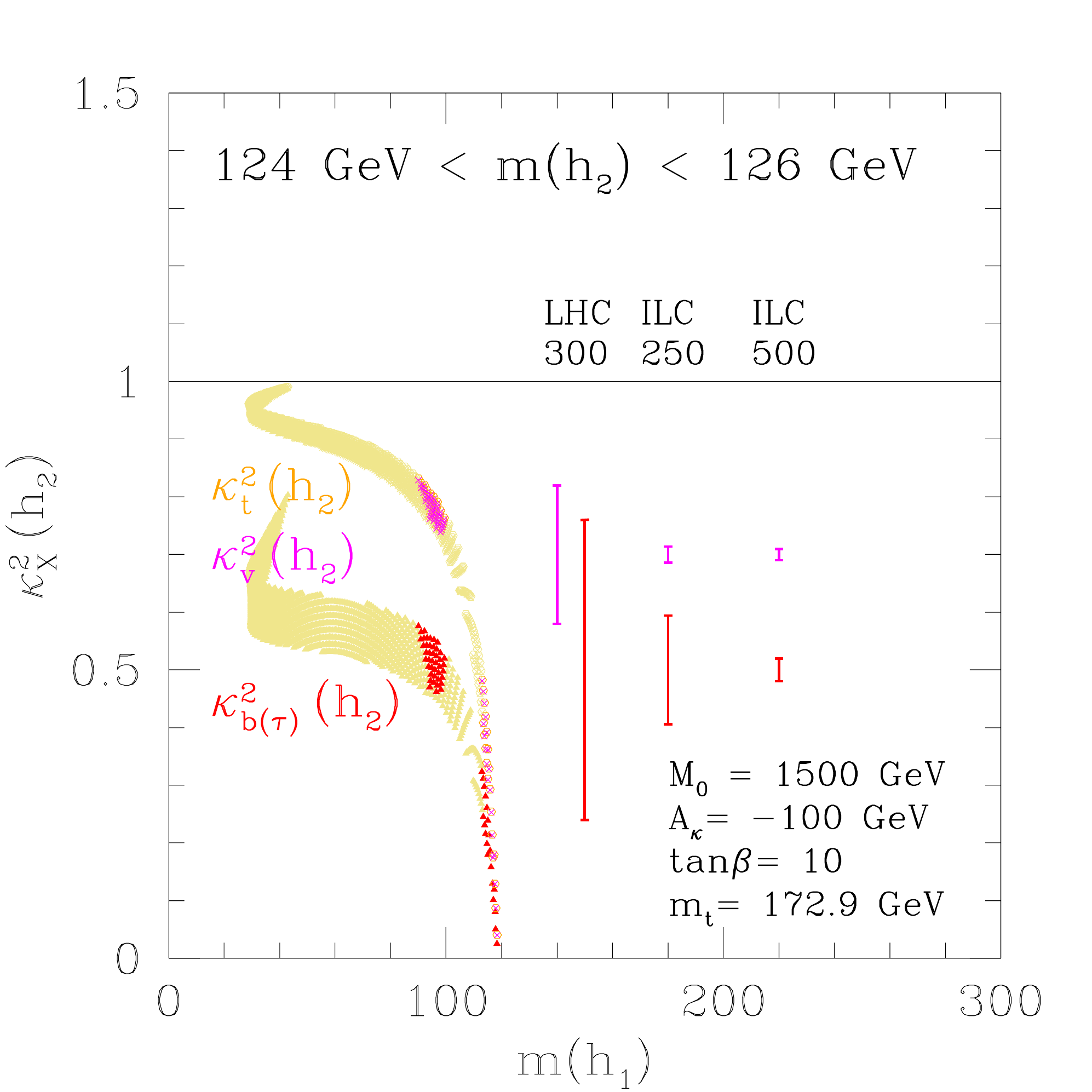} &
\includegraphics[height=60mm]{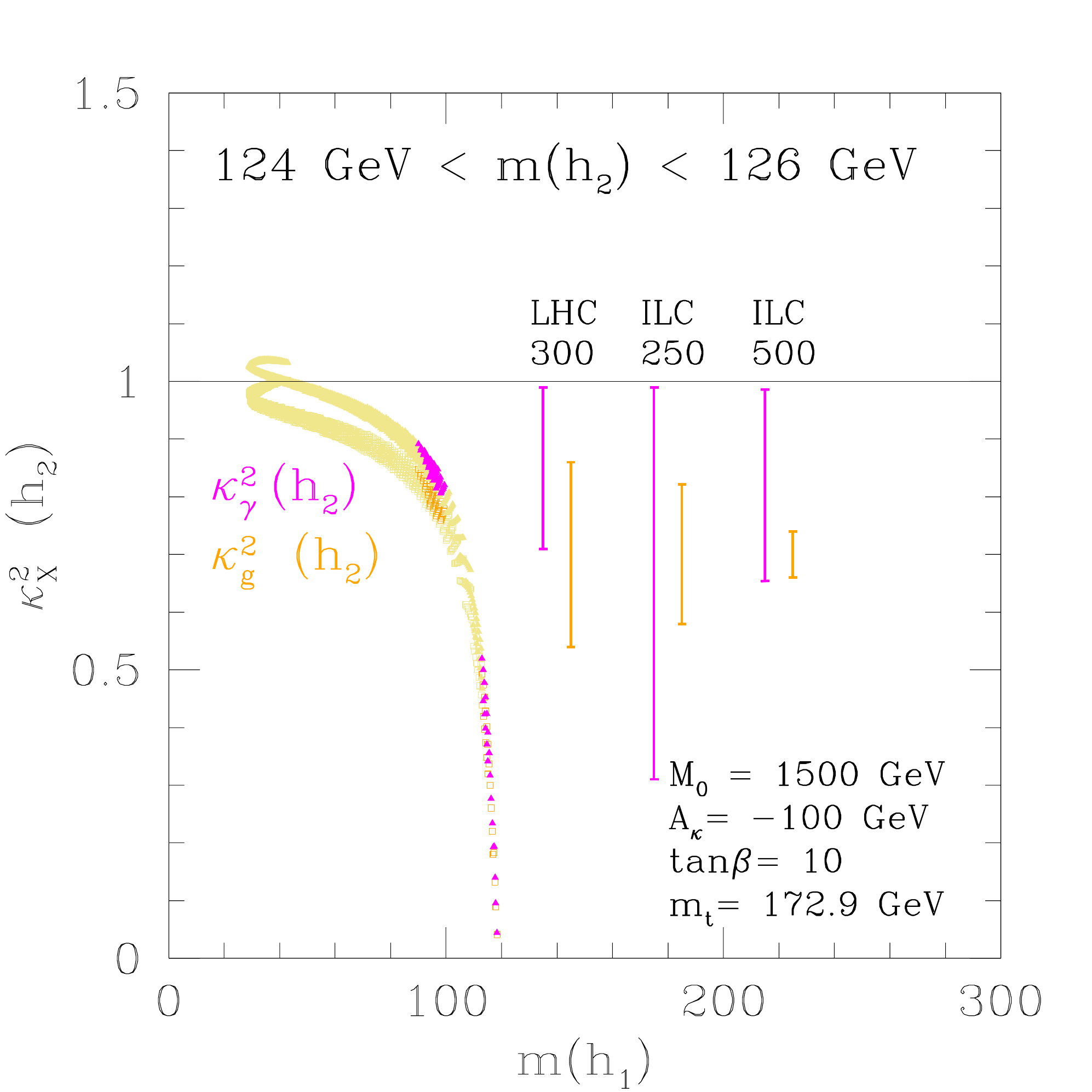} \\ 
\end{tabular}

\caption{\label{fig:cpevencouplings-tanb10}The scale factors of the CP-even Higgs coupling constants for $\tan\beta=10$, $M_0=1500$ GeV and $A_\kappa=-100$ GeV.
The vertical lines show the expected precision of the SM-like Higgs coupling constants for LHC and ILC, $\Delta (\kappa_X)^2 \approx 2 \Delta \kappa_X$.
$(\kappa_V^1)^2$ and $(\kappa_t^1)^2$ ($(\kappa_V^2)^2$ and $(\kappa_t^2)^2$) in the upper (lower) left panel almost overlap each other.
}
\end{center}
\end{figure}

In figure \ref{fig:cpevencouplings-tanb10}, we plot the square of the scaling factors for $\tan\beta=10$. The other parameters are same as in figure \ref{fig:cpevencouplings-tanb3}. Now the second lightest CP even boson is the SM like.
In the figure, the khaki regions are phenomenologically excluded mainly by the LEP Higgs boson search. Remember that the surviving region has the lowest fine-tuning as we saw in the previous section.
We can see the sum rule holds well again except for the bottom quark and tau lepton.
The deviations of the scale factors for the SM-like Higgs boson amount to 10\% (30\%) for the vector bosons, top quark and loop induced couplings (the bottom quark and the tau lepton). 
The enhancement of the deviations occurs because the SM-like Higgs boson requires a sizable mixing with the lighter singlet to achieve 125 GeV, since $\lambda$ is now not effective to raise the Higgs boson mass.
If we increase $M_0$, the surviving region extends while the deviations shrink since the mixing is replaced with the top radiative correction to lift the SM-like Higgs boson mass. Simultaneously, the fine-tuning deteriorates as in figure \ref{fig:heavy-case}.
In figure \ref{fig:cpevencouplings-tanb10}, The couplings of the singlet-like Higgs boson to the bottom quark and the tau lepton can be 80\% of those for the SM Higgs boson due to $\tan\beta$ enhanced contribution of the heavy doublet, while those to the vector bosons and top quark is 40\% of the SM.
It is hard to examine this low energy region by LHC.
ILC has more advantage to rediscover the particles escaped from LEP. 

\section{Conclusion}

We proposed that the TeV scale mirage mediation in NMSSM is a novel solution for the little hierarchy problem in SUSY. 
The hierarchy between the electroweak scale and the SUSY scale is realized by the mirage unification.
The SM-like Higgs boson reaches 125 GeV by the new quartic coupling or the mixing with the singlet.
We calculated the fine-tuning measures of the electroweak symmetry breaking in the model. 
The fine-tuning is better than 10\% with $1.5$ TeV gluino and $1$ TeV stop, while the effective $\mu$ can be as heavy as $500$ GeV thanks to a cancellation mechanism.
An undesirable singlet-doublet mixing is suppressed due to the approximate scale symmetries.
We found $5$-$10$\% deviation in the SM-like Higgs couplings.
${\mathcal O}(1)$ deviation is possible for the bottom quark and the tau lepton due to $\tan\beta$ enhanced couplings of the heavy doublet and its mixing in the SM-like Higgs boson.
The singlet-like Higgs boson is around the electroweak scale and its couplings are suppressed due to the scale symmetries. It is an interesting target for future lepton colliders.


\begin{thebibliography}{99}

\bibitem{Hagimoto:2015tua}
  K.~Hagimoto, T.~Kobayashi, H.~Makino, K.~i.~Okumura and T.~Shimomura,
  JHEP {\bf 1602} (2016) 089
  doi:10.1007/JHEP02(2016)089
  [arXiv:1509.05327 [hep-ph]].


\bibitem{Aad:2012tfa}
  G.~Aad {\it et al.}  [ATLAS Collaboration],
  Phys.\ Lett.\ B {\bf 716} (2012) 1
  [arXiv:1207.7214 [hep-ex]].
\bibitem{Chatrchyan:2012ufa}
  S.~Chatrchyan {\it et al.}  [CMS Collaboration],
  Phys.\ Lett.\ B {\bf 716} (2012) 30
  [arXiv:1207.7235 [hep-ex]].

\bibitem{Aad:2015zhl}
  G.~Aad {\it et al.}  [ATLAS and CMS Collaborations],
  arXiv:1503.07589 [hep-ex].

\bibitem{Ulmer:2016ljv}
  K.~A.~Ulmer [CMS and ATLAS Collaborations],
  arXiv:1601.03774 [hep-ex].

\bibitem{Choi:2005hd}
K.~Choi, K.~S.~Jeong, T.~Kobayashi and K.~i.~Okumura,
Phys.\ Lett.\  B {\bf 633}, 355 (2006)
[arXiv:hep-ph/0508029].

\bibitem{Choi:2006xb}
K.~Choi, K.~S.~Jeong, T.~Kobayashi and K.~i.~Okumura,
Phys.\ Rev.\  D {\bf 75}, 095012 (2007)
[arXiv:hep-ph/0612258].


\bibitem{Kitano:2005wc}
R.~Kitano and Y.~Nomura,
Phys.\ Lett.\  B {\bf 631}, 58 (2005)
[arXiv:hep-ph/0509039].

\bibitem{Kobayashi:2012ee} 
  T.~Kobayashi, H.~Makino, K.~i.~Okumura, T.~Shimomura and T.~Takahashi,
JHEP {\bf 1301}, 081 (2013)  [arXiv:1204.3561 [hep-ph]].  


\bibitem{Asano:2012sv} 
M.~Asano and T.~Higaki,
Phys.\ Rev.\ D {\bf 86}, 035020 (2012)  [arXiv:1204.0508 [hep-ph]].  


\bibitem{Choi:2004sx}
K.~Choi, A.~Falkowski, H.~P.~Nilles, M.~Olechowski and S.~Pokorski,
JHEP {\bf 0411}, 076 (2004)
[arXiv:hep-th/0411066];
%
%
K.~Choi, A.~Falkowski, H.~P.~Nilles and M.~Olechowski,
Nucl.\ Phys.\  B {\bf 718}, 113 (2005)
[arXiv:hep-th/0503216].

\bibitem{Choi:2005uz}
K.~Choi, K.~S.~Jeong and K.~i.~Okumura,
JHEP {\bf 0509}, 039 (2005)
[arXiv:hep-ph/0504037].

\bibitem{Endo:2005uy}
M.~Endo, M.~Yamaguchi and K.~Yoshioka,
Phys.\ Rev.\  D {\bf 72}, 015004 (2005)
[arXiv:hep-ph/0504036].

\bibitem{Kachru:2003aw}
  S.~Kachru, R.~Kallosh, A.~D.~Linde and S.~P.~Trivedi,
  Phys.\ Rev.\ D {\bf 68} (2003) 046005
  doi:10.1103/PhysRevD.68.046005
  [hep-th/0301240].


\bibitem{Kaplunovsky:1993rd}
V.~S.~Kaplunovsky and J.~Louis,
Phys.\ Lett.\  B {\bf 306}, 269 (1993)
[arXiv:hep-th/9303040];
%
%
%
A.~Brignole, L.~E.~Ibanez and C.~Munoz,
Nucl.\ Phys.\  B {\bf 422}, 125 (1994)
[Erratum-ibid.\  B {\bf 436}, 747 (1995)]
[arXiv:hep-ph/9308271];
%
T.~Kobayashi, D.~Suematsu, K.~Yamada and Y.~Yamagishi,
Phys.\ Lett.\  B {\bf 348}, 402 (1995)
[arXiv:hep-ph/9408322];
%
L.~E.~Ibanez, C.~Munoz and S.~Rigolin,
Nucl.\ Phys.\  B {\bf 553}, 43 (1999)
[arXiv:hep-ph/9812397].

\bibitem{Randall:1998uk}
L.~Randall and R.~Sundrum,
Nucl.\ Phys.\  B {\bf 557}, 79 (1999)
[arXiv:hep-th/9810155];
%
G.~F.~Giudice, M.~A.~Luty, H.~Murayama and R.~Rattazzi,
JHEP {\bf 9812}, 027 (1998)
[arXiv:hep-ph/9810442].


\bibitem{Choi:2008hn} 
K.~Choi, K.~S.~Jeong and K.~i.~Okumura,
JHEP {\bf 0807}, 047 (2008)  [arXiv:0804.4283 [hep-ph]].  


\bibitem{Nakamura:2008ey}
  S.~Nakamura, K.~i.~Okumura and M.~Yamaguchi,
  Phys.\ Rev.\ D {\bf 77} (2008) 115027
  [arXiv:0803.3725 [hep-ph]].


\bibitem{Fayet:1974pd}
P.~Fayet,
Nucl.\ Phys.\  B {\bf 90}, 104 (1975);
%
Phys.\ Lett.\  B {\bf 64}, 159 (1976);
%
Phys.\ Lett.\  B {\bf 69}, 489 (1977);
%
Phys.\ Lett.\  B {\bf 84}, 416 (1979);
%
H.~P.~Nilles, M.~Srednicki and D.~Wyler,
Phys.\ Lett.\  B {\bf 120}, 346 (1983);
%
J.~M.~Frere, D.~R.~T.~Jones and S.~Raby,
Nucl.\ Phys.\  B {\bf 222}, 11 (1983);
%
J.~P.~Derendinger and C.~A.~Savoy,
Nucl.\ Phys.\  B {\bf 237}, 307 (1984);
%
J.~R.~Ellis, J.~F.~Gunion, H.~E.~Haber, L.~Roszkowski and F.~Zwirner,
Phys.\ Rev.\  D {\bf 39}, 844 (1989);
%
M.~Drees,
Int.\ J.\ Mod.\ Phys.\  A {\bf 4}, 3635 (1989).

\bibitem{Ellwanger:2009dp}
U.~Ellwanger, C.~Hugonie and A.~M.~Teixeira,
Phys.\ Rept.\  {\bf 496}, 1 (2010)
[arXiv:0910.1785 [hep-ph]].

\bibitem{Maniatis:2009re}
  M.~Maniatis,
  Int.\ J.\ Mod.\ Phys.\ A {\bf 25} (2010) 3505
  [arXiv:0906.0777 [hep-ph]].

\bibitem{Hattori:2015xla}
  H.~Hattori, T.~Kobayashi, N.~Omoto and O.~Seto,
  Phys.\ Rev.\ D {\bf 92} (2015) no.10,  103518
  doi:10.1103/PhysRevD.92.103518
  [arXiv:1510.03595 [hep-ph]].



\bibitem{Kim:1983dt}
J.~E.~Kim and H.~P.~Nilles,
Phys.\ Lett.\  B {\bf 138}, 150 (1984).

\bibitem{natural-SUSY}
  M.~Papucci, J.~T.~Ruderman and A.~Weiler,
  JHEP {\bf 1209} (2012) 035
  [arXiv:1110.6926 [hep-ph]];
  H.~Baer, V.~Barger, P.~Huang and X.~Tata,
  JHEP {\bf 1205} (2012) 109
  [arXiv:1203.5539 [hep-ph]];
  H.~Baer, V.~Barger, P.~Huang, A.~Mustafayev and X.~Tata,
  Phys.\ Rev.\ Lett.\  {\bf 109} (2012) 161802
  [arXiv:1207.3343 [hep-ph]];
  H.~Baer, V.~Barger, P.~Huang, D.~Mickelson, A.~Mustafayev and X.~Tata,
  Phys.\ Rev.\ D {\bf 87} (2013) 11,  115028
  [arXiv:1212.2655 [hep-ph]];
  H.~Baer, V.~Barger, P.~Huang, D.~Mickelson, A.~Mustafayev, W.~Sreethawong and X.~Tata,
  JHEP {\bf 1312} (2013) 013
  [arXiv:1310.4858 [hep-ph]];
  O.~Buchmueller and J.~Marrouche,
  Int.\ J.\ Mod.\ Phys.\ A {\bf 29} (2014) 06,  1450032
  [arXiv:1304.2185 [hep-ph]];
  G.~D.~Kribs, A.~Martin and A.~Menon,
  Phys.\ Rev.\ D {\bf 88} (2013) 035025
  [arXiv:1305.1313 [hep-ph]];
  J.~A.~Evans, Y.~Kats, D.~Shih and M.~J.~Strassler,
  JHEP {\bf 1407} (2014) 101
  [arXiv:1310.5758 [hep-ph]];
  S.~P.~Martin,
  Phys.\ Rev.\ D {\bf 89} (2014) 3,  035011
  [arXiv:1312.0582 [hep-ph]].



\bibitem{Ellwanger:2005dv} 
 U.~Ellwanger and C.~Hugonie,
 Comput.\ Phys.\ Commun.\  {\bf 175}, 290 (2006)
 [hep-ph/0508022];
 G.~Belanger, F.~Boudjema, C.~Hugonie, A.~Pukhov and A.~Semenov,
 JCAP {\bf 0509}, 001 (2005)
 [hep-ph/0505142];
 U.~Ellwanger, J.~F.~Gunion and C.~Hugonie,
 JHEP {\bf 0502}, 066 (2005)
 [hep-ph/0406215].


\bibitem{Kanehata:2011ei}
Y.~Kanehata, T.~Kobayashi, Y.~Konishi, O.~Seto and T.~Shimomura,
Prog.\ Theor.\ Phys.\  {\bf 126}, 1051 (2011)
[arXiv:1103.5109 [hep-ph]];
%
  T.~Kobayashi, T.~Shimomura and T.~Takahashi,
Phys.\ Rev.\ D {\bf 86}, 015029 (2012)  [arXiv:1203.4328 [hep-ph]].  


\bibitem{FT}
  J.~R.~Ellis, K.~Enqvist, D.~V.~Nanopoulos and F.~Zwirner,
  Mod.\ Phys.\ Lett.\ A {\bf 1}, 57 (1986);
  R.~Barbieri and G.~F.~Giudice,
  Nucl.\ Phys.\ B {\bf 306} (1988) 63.

\bibitem{Dimopoulos:1995mi} 
  S.~Dimopoulos and G.~F.~Giudice,
  Phys.\ Lett.\ B {\bf 357}, 573 (1995)
  [hep-ph/9507282].

\bibitem{FTLEP}
  P.~H.~Chankowski, J.~R.~Ellis and S.~Pokorski,
  Phys.\ Lett.\ B {\bf 423}, 327 (1998)
  [hep-ph/9712234];
  P.~H.~Chankowski, J.~R.~Ellis, M.~Olechowski and S.~Pokorski,
  Nucl.\ Phys.\ B {\bf 544}, 39 (1999)
  [hep-ph/9808275];
  R.~Barbieri and A.~Strumia,
  Phys.\ Lett.\ B {\bf 433}, 63 (1998)
  [hep-ph/9801353];
  G.~L.~Kane and S.~F.~King,
  Phys.\ Lett.\ B {\bf 451}, 113 (1999)
  [hep-ph/9810374];
  L.~Giusti, A.~Romanino and A.~Strumia,
  Nucl.\ Phys.\ B {\bf 550}, 3 (1999)
  [hep-ph/9811386].



\bibitem{Schael:2006cr}
  S.~Schael {\it et al.}  [ALEPH and DELPHI and L3 and OPAL and LEP Working Group for Higgs Boson Searches Collaborations],
  Eur.\ Phys.\ J.\ C {\bf 47} (2006) 547
  [hep-ex/0602042].


\bibitem{Belanger:2012tt}
  G.~Belanger, U.~Ellwanger, J.~F.~Gunion, Y.~Jiang, S.~Kraml and J.~H.~Schwarz,
  JHEP {\bf 1301} (2013) 069
  [arXiv:1210.1976 [hep-ph]];
  D.~G.~Cerdeno, P.~Ghosh and C.~B.~Park,
  JHEP {\bf 1306} (2013) 031
  [arXiv:1301.1325 [hep-ph]];
  B.~Bhattacherjee, M.~Chakraborti, A.~Chakraborty, U.~Chattopadhyay, D.~Das and D.~K.~Ghosh,
  Phys.\ Rev.\ D {\bf 88} (2013) 3,  035011
  [arXiv:1305.4020 [hep-ph]];
  R.~Barbieri, D.~Buttazzo, K.~Kannike, F.~Sala and A.~Tesi,
  Phys.\ Rev.\ D {\bf 88} (2013) 055011
  [arXiv:1307.4937 [hep-ph]].

\bibitem{Drees:2005jg}
  M.~Drees,
  Phys.\ Rev.\ D {\bf 71} (2005) 115006
  [hep-ph/0502075],
  Phys.\ Rev.\ D {\bf 86} (2012) 115018
  [arXiv:1210.6507 [hep-ph]].

\bibitem{Dermisek:2005gg}
  R.~Dermisek and J.~F.~Gunion,
  Phys.\ Rev.\ D {\bf 73} (2006) 111701
  [hep-ph/0510322],
  Phys.\ Rev.\ D {\bf 76} (2007) 095006
  [arXiv:0705.4387 [hep-ph]].

\bibitem{Carmi:2012in}
  D.~Carmi, A.~Falkowski, E.~Kuflik, T.~Volansky and J.~Zupan,
  JHEP {\bf 1210} (2012) 196
  [arXiv:1207.1718 [hep-ph]].


\bibitem{LHCHiggsCrossSectionWorkingGroup:2012nn}
  A.~David {\it et al.} [LHC Higgs Cross Section Working Group Collaboration],
  arXiv:1209.0040 [hep-ph].
\bibitem{Heinemeyer:2013tqa}
  S.~Heinemeyer {\it et al.} [LHC Higgs Cross Section Working Group Collaboration],
  arXiv:1307.1347 [hep-ph].

\bibitem{Dawson:2013bba}
  S.~Dawson, A.~Gritsan, H.~Logan, J.~Qian, C.~Tully, R.~Van Kooten, A.~Ajaib and A.~Anastassov {\it et al.},
  arXiv:1310.8361 [hep-ex].

\bibitem{Asner:2013psa}
  D.~M.~Asner, T.~Barklow, C.~Calancha, K.~Fujii, N.~Graf, H.~E.~Haber, A.~Ishikawa and S.~Kanemura {\it et al.},
  arXiv:1310.0763 [hep-ph].


\end{thebibliography}
\end{document}